\documentstyle{mn}

\input epsf

\def\et{et\thinspace al.\ }                                 

\def\kms{km\thinspace s$^{-1}$}                        
\def\Lsun{L$_{\odot}$}                            

\newcommand{\ET}[1]{\times 10^{#1}}             

\title[Patterns in the Light Curves of Active Galaxies]
      {Patterns and Coincidences in the Light Curves of Active
Galaxies}

\author[R. Cid Fernandes, R. Terlevich, I. Aretxaga]
       {Roberto Cid Fernandes Jr.$^{1\mbox{$\dagger$}}$,
        Roberto Terlevich$^{2\mbox{$\dagger$}}$, and
        Itziar Aretxaga$^2$\thanks{Present address: Max Planck 
Institut f\"ur Astrophysik, Karl Schwarzschild str.\ 1, 85740 Garching, 
Germany}\thanks{E-mail: cid@fsc.ufsc.br (RCF); rjt@ast.cam.ac.uk (RT); 
\newline itziar@mpa-garching.mpg.de (IA).}
       \\
       $^1$ Departamento de F\'{\i}sica, CFM, UFSC. Campus
Universit\'ario, Trindade, Caixa Postal 476, 88040-900 Florian\'opolis,
SC, Brazil \\
       $^2$ Royal Greenwich Observatory, Madingley Road, CB3~0EZ,
Cambridge}

\begin{document}

\maketitle

    \begin{abstract}

    We draw attention to some intriguing coincident patterns in the
optical light curves of NGC 4151 and NGC 5548. Inspecting the light
curves of these two classic Seyferts, we find evidence for at least
five occurrences of a similar sequence of variations. This pattern
consists of a $\sim 2$~yr long sequence of shorter flares, with a
total energy of $3$--$5\ET{50}$~ergs in the B-band. We predict that
the on-going monitoring campain on NGC 5548 should soon reveal more
occurrences of this general pattern if it is truly recurrent. We
speculate on the possibility that this pattern is associated with a
fundamental/universal `unit' of variability in active galaxies, and
discuss a possible connection with the variability properties of
QSOs.

    \end{abstract}


\section{Introduction}

Optical--UV variability was long ago established as a fundamental
property of active galactic nuclei (AGN) and has riddled us ever
since. In recent times astronomers have focused their interest not
so much on the physical origin of the continuum variations, but into
using them to infer the geometry and dynamics of the emitting gas of
the broad line region through reverberation mapping techniques
(e.g.\ Peterson 1993 and references therein). Though emission line
variability studies have witnessed a substantial progress in the
past few years, the nature of the continuum variations themselves is
still not well understood.

AGN vary in a rather chaotic way, with no indications of
periodicities, but some regularities have been reported. In the
seventies, Lyutyi and co-workers suggested that AGN variability
could be described in terms of two basic components: (1) a {\it
rapid component}: a sharp peak with a time scale of weeks, and (2) a
{\it slow component}: a broad bump a few years long, on top of which
the rapid component sits (Lyutyi 1972, 1977, 1979, Diba\u{\i}\ \&
Lyutyi 1984, Lyutyi \& Oknyanskii 1987). This overall pattern was
further confirmed by Gill \et (1984), Pica \& Smith (1983), Alloin
\et (1986) and Smith \et (1993). Alloin and co-workers, in
particular, found that NGC 1566 exhibits recurrent outbursts
interspaced by quiescent periods. The global time scale of the
outbursts is $\sim 3.5$~yr, but the brightest stages last only a few
weeks. 

One of the major goals of variability studies is, of course, to
identify and characterize the physical processes responsible for the
variations. Establishing whether recurrent patterns or well defined
sequences of events occur in the light curves of AGN would be an
important step towards this goal. This paper re-opens the discussion
of variability patterns by examining some intriguing coincidences in
the optical light curves of the two best monitored type 1 Seyferts:
NGC 4151 and NGC 5548. Section~\ref{sec:DATA} introduces the data
sets used in this work. In Section~\ref{sec:PULSES} we identify
several occurrences of the same pattern of variations in these two
light curves. The significance of these coincidences is assessed in
Section~\ref{sec:XCORREL} by means of cross-correlation techniques. 
A discussion of our results, their potential implications for AGN
models and a possible extrapolation to high luminosity QSOs is
presented in Section~\ref{sec:DISCUSSION}. Finally,
Section~\ref{sec:CONCLUSIONS} summarizes our findings.

\section{The optical light curves of NGC 4151 and NGC 5548}

\label{sec:DATA}

The data used in this study are the B-band light curves of NGC 4151
and NGC 5548. The data for NGC 4151 consist of 834 observations in
the 1968--1990 period, compiled and intercalibrated by M. Penston
and T. Snijders, who also subtracted the contribution from the host
galaxy (Snijders 1991---see also Gill \et 1984 and Lyutyi \&
Oknyanskii 1987). The apparent magnitudes were converted to  B-band
luminosity ($L_B$) applying a $E_{B-V} = 0.16$\ reddening correction
(Rieke \& Lebofsky 1981) and assuming a distance of $27$~Mpc. This
distance includes a correction to allow for the infall of NGC 4151
towards Virgo (Aretxaga \& Terlevich 1994).

NGC 5548 has been intensely monitored by the AGN Watch consortium
since 1989 (Korista \et 1995 and references therein). The data used
here consist of monochromatic fluxes at 5100\AA\ for 559 points
covering the 4.8~yr interval between 1988.9 and 1993.7. The
estimated host galaxy flux in the $7.5'' \times 5''$\ slit used in
the observations is
$3.4\ET{-15}$~ergs$\,$s$^{-1}$cm$^{-2}$\AA$^{-1}$\ at $5100$\ \AA\
(Korista \et 1995, Peterson \et 1995, Romanishin \et 1995). Once
this component was subtracted, the nuclear fluxes at 5100\AA\ were
converted to B-band ($3900$--$4900$~\AA) assuming a spectral index
$\alpha = -1.23$\ ($F_\nu \propto \nu^\alpha$, Edelson \& Malkan
1986). $L_B$\ was then computed applying a $E_{B-V} = 0.1$\
reddening correction (Tsvetanov \& Yancoulova 1989) and assuming a
distance of $103$~Mpc, corresponding to $z = 0.0173$ (de Vaucouleurs
\et 1991) and $H_0 = 50$~\kms\,Mpc$^{-1}$.

\begin{figure}  
    \protect\centerline{
    \epsfxsize=8.6cm\epsffile[65 190 305 695]{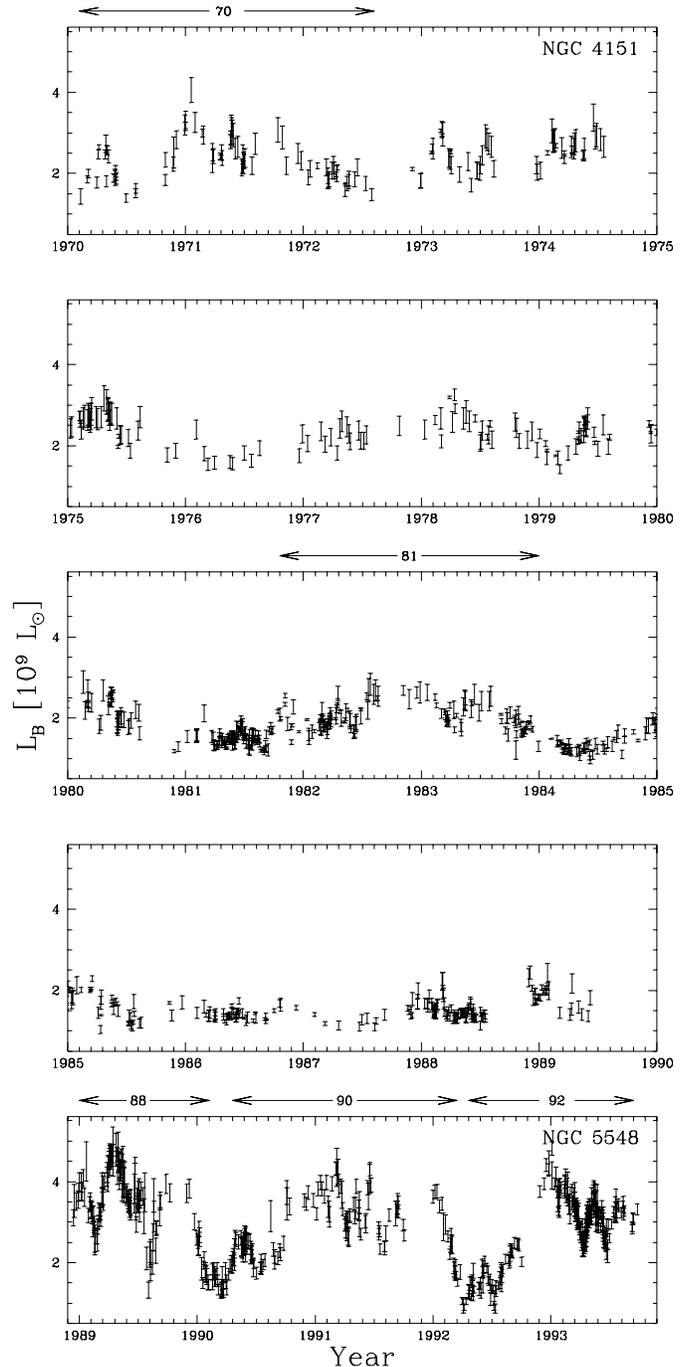}
     }
    \caption{
    B-band light curves of NGC 4151 and NGC 5548. Pulses 70, 81, 88,
90 and 92 discussed in the text are indicated.
    }
    \label{fig:LCs}
\end{figure}

The two light curves are plotted in Fig.~\ref{fig:LCs}. The
peak-to-peak variations in the light curves of these two type 1
Seyferts are $3\ET{9}$~\Lsun\ for NGC 4151 and $4\ET{9}$~\Lsun\ for
NGC 5548. The minimum recorded luminosities for these two objects
are also similar: $10^9$~\Lsun\ for NGC 4151 (which ocurred in
1984.4) and $9\ET{8}$~\Lsun\ for NGC 5548 (in 1992.5). Taking into
account the uncertainties in the reddening correction, starlight
subtraction and distances, these values can be regarded as
consistent with each other. In any case, the identification of
patterns in the variations is not critically affected by the
uncertainties in the absolute luminosity scales.

\section{Identification of Patterns}

\label{sec:PULSES}

The identification of patterns in the data is complicated by the
numerous gaps in the observations and the less than ideal error
bars. Yet, a close examination of Fig.~\ref{fig:LCs} reveals some
surprising coincidences.

\subsection{Pulse 92 in NGC 5548 and Pulse 70 in NGC 4151}

\begin{figure}  
    \protect\centerline{
    \epsfxsize=8.5cm\epsffile[60 450 360 705]{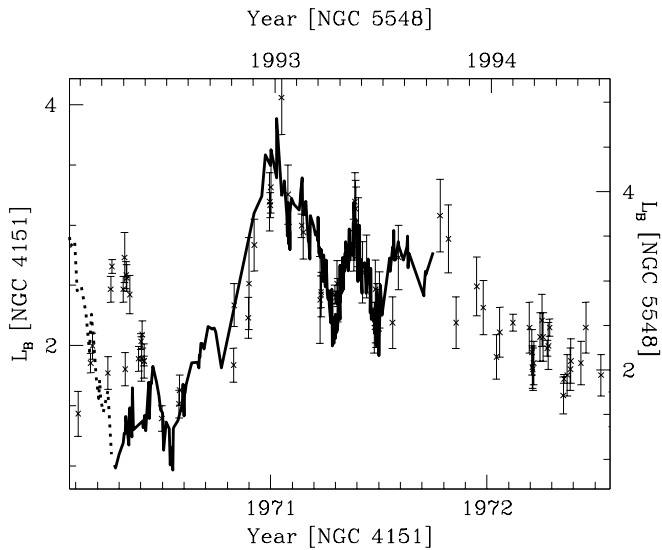}
     }
    \caption{
    Superposition of the light curves of NGC 4151 from 1970 to 1973
(error bars) to that of NGC 5548 twenty two years later (thick
dotted and solid lines). Bottom and left axes correspond to NGC 4151
while top and right axes to NGC 5548. Luminosities in units of
$10^9$~\Lsun.
    }
    \label{fig:70X92}
\end{figure}

\begin{figure*}  
\begin{minipage}{17.5cm}
    \protect\centerline{
    \epsfxsize=17.5cm\epsffile[35 260 565 705]{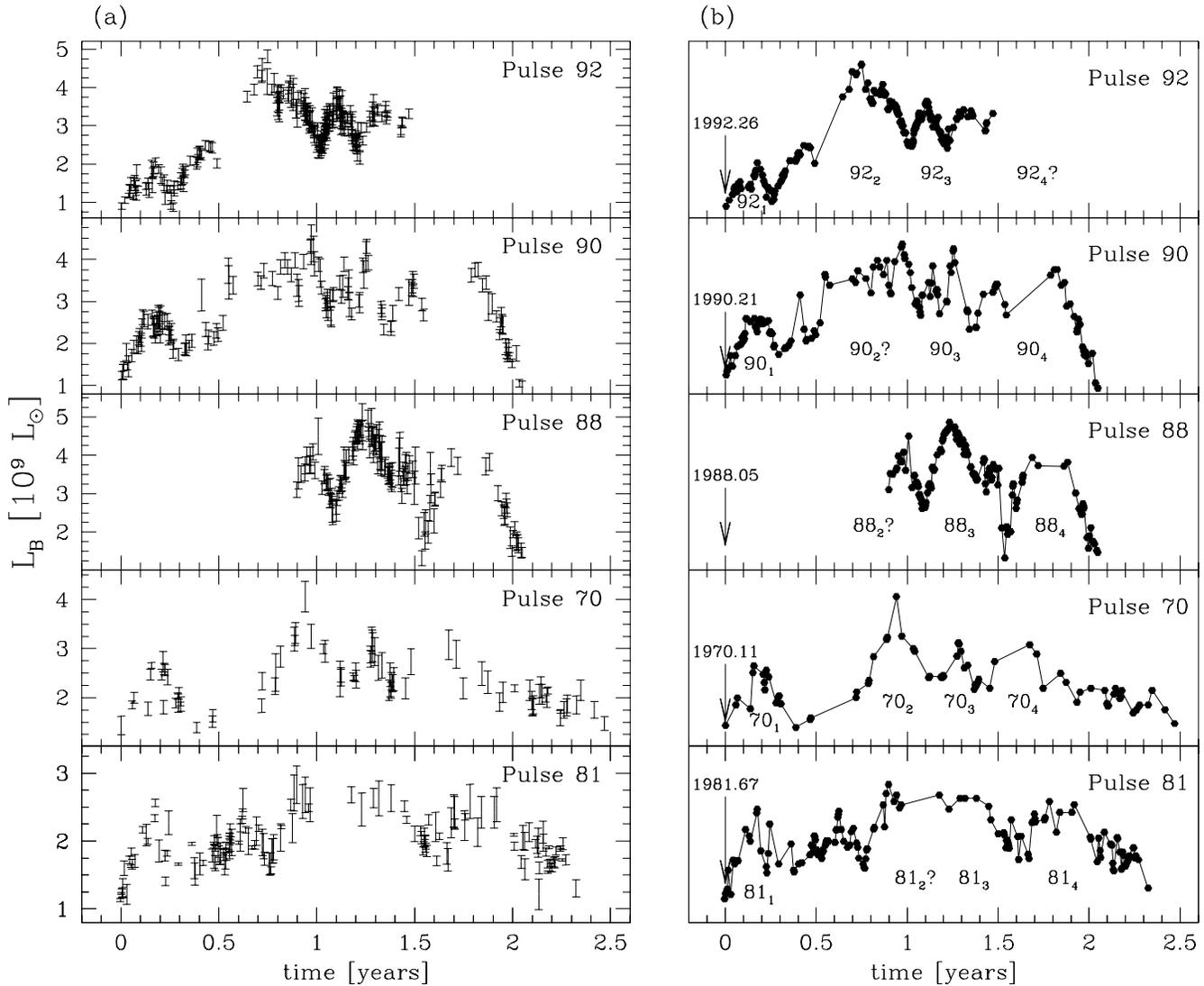}
     }
    \caption{
    (a) Blow-up of pulses 70, 81, 88, 90 and 92. (b) Same as in
panel a, but after smoothing the pulses with a Gaussian filter with
a standard deviation of 1 day. The starting dates are indicated.
Dots correspond to observed dates.
    }
    \label{fig:Pulses}
\end{minipage}
\end{figure*}

The first similarity which strikes the eye when comparing the two
light curves in Fig.~\ref{fig:LCs} is that between the `pulse'
starting in 1992.3 and going till the end of the data train at
1993.7 in NGC 5548 (hereafter referred to as `pulse 92') and the
variations of NGC 4151 between 1970.1 and 1972.6 (`pulse 70'). The
two sequences are overlayed in Fig.~\ref{fig:70X92}. Pulses 70, 92
and three other similar events discussed below are plotted in
Fig.~\ref{fig:Pulses}.

Both pulses 70 and 92 start with an initial rapid, $\sim 0.3$~yr
long flare. These flares are labeled $70_1$\ and $92_1$\ in
Fig.~\ref{fig:Pulses}b, where in order to better illustrate the main
features in the light curves we have smoothed the pulses by a
Gaussian filter with a standard deviation of 1 day (the original
data is show in Fig.~\ref{fig:Pulses}a). This rapid flare seems to
play the role of a `starting gun' to the big and long lasting flare
($70_2$--$70_4$\ and $92_2$--$92_3$) which follows it. In NGC 4151
the second and most luminous peak ($70_2$) occurs $0.75$~yr after
the peak of the starting shot, while the corresponding delay in NGC
5548 is $0.6$~yr ($92_2$). The amplitudes and timing of the starting
shots are somewhat different in the two objects, but from then on
their light curves are remarkably similar. After the maximum
luminosity both objects undergo a rapid decay, which is reversed
$\sim 0.25$~yr after the maximum, leading to a new peak ($70_3$\ and
$92_3$) $\sim 0.1$~yr later. The overall pattern of the `slow
component' could be described as that of a `damped oscillator': a
decaying pulse with a time-scale of $\sim 1$--$2$~yr with $\sim
0.3$~yr oscillations on top of it.

Pulses 70 and 92 may be somewhat different quantitatively (note the
different scales for the two pulses in Fig.~\ref{fig:70X92}), but
are similar qualitatively. This similarity strongly suggests a
common underlying physical process causing the variations. The most
direct way of assessing the reality of this coincidence is to look
for other occurrences of this general pattern. At least three other
such coincidences can be found in the data. These are discussed
next. 

\subsection{Pulses 88, 90 and 92 in NGC 5548}

We now compare pulse 92 in NGC 5548 with pulse 90, defined as the
interval  between 1990.2 and 1992.2 also in NGC 5548. Pulses 92 and
90 have similar starting shots ($92_1$\ and $90_1$) and initial rise
after that. The second peak ($92_2$), however, either was not as
strong in 90 as in 92 or it was simply missed by the sparse sampling
in this region of pulse 90. The local minimum at 1991.3 in 90,
however, is well matched by that later observed in 1993.3 in pulse
92. Overall, the amplitude and time-scale of the two pulses match
each other reasonably well. Note that pulse 92 is incomplete, so the
late behavior of the pulses cannot be compared. Unless a new pulse
occurred late in 1993, we expect the NGC 5548 light curve to present
a broad hump as that observed between 1991.6 and 1992.2 in pulse 90
($90_4$). This prediction can be readily tested when the AGN Watch
group publishes the data for the period after 1993.7. In fact, since
nearly four years have gone by since the last observation of NGC
5548 used in this paper, the new data will be extremely important to
test whether patterns such as pulses 90 and 92 are really recurrent.

The variations in the period from 1989 to 1990 in NGC 5548 (`pulse
88') also seem to fit on a pattern similar to that of pulses 90 and
92. This is best seen shifting pulse 90 back by 2 years, so that it
starts around 1988.1 (see Fig.~\ref{fig:Pulses}). Though the initial
phases of the pulse evolution cannot be compared  given that the
observations started in the `middle' of the pulse, there is a good
correspondance between the peaks at 1991.2 and 1989.0, and between
the peaks labeled $90_4$\ and $88_4$\ in Fig.~\ref{fig:Pulses}b. The
only substantial difference between these two pulses is that while
pulse 88 has a single $0.2$~yr long peak when the pulse is $\sim
1.1$~yr old (at 1989.3), the corresponding phase in pulse 90 shows
two smaller peaks (see Fig.~\ref{fig:Pulses}). Overall, however, the
matching is good in both amplitude and time scale.

\subsection{Pulses 70 and 81 in NGC 4151}

\label{sec:70X81}

Going back to the NGC 4151 light curve, an event similar to pulse 70
was observed in the period from 1981.7 to 1984.0 (`pulse 81'). The
starting shots ($70_1$\ and $81_1$) have similar amplitudes and
durations. Comparing the two pulses we see that the global maximum
(at 1971.05 in pulse 70) was probably missed in pulse 81---it should
have occurred around 1982.6, where there is a gap in the data. The
hump at 1983.0 ($81_3$) does match that of $70$\ at 1971.4 ($70_3$),
as does the later decay in luminosity. The agreement here is not as
impressive as that between 70 and 92 or between 92, 90 and 88, but
the global behavior of the two pulses is undoubtedly similar.

\subsection{Other possible coincidences in NGC 4151}

\label{sec:Other_Pulses}

A few other possible matches can be found in the light curve of NGC
4151 sliding pulse 70 (or pulse 81)  over the light curve so that it
starts at any of the following dates: 1972.3, 1973.5, 1976.9 and
1978.6 (see also Section~\ref{sec:XCORREL} below). None of these
matches are as convincing as the previous ones, but it must be
stressed that this sort of analysis is severely affected by the poor
sampling throughout most of the light curve of NGC 4151. Also,
pulses may ocasionally overlap in time, which would complicate their
identification. Such a superposition of pulses does indeed seem to
have occurred in NGC 4151 between $\sim$\ 1973.5 and 1974.5. If we
associate the small $\sim 0.3$~yr long flare centered on 1972.5 to a
starting shot like $70_1$, then the 1973.2 and 1973.55 peaks are
very well matched by $70_2$\ and $70_3$\ respectively. From 1974
onwards, however, the light curve reverses its downward trend and
goes into a new luminous phase which lasts until the end of 1975.
This indicates that a new pulse occurred sometime around 1973.5,
probably overlapping with the $70_3$-like peak of the previous
event. The starting shot of this new event would have peaked around
1973.7, where there is a gap in the data. The second and third
maxima, corresponding to $70_2$\ and $70_3$, would be placed at
$\sim$\ 1974.5, where a local maximum is indeed found, and 1974.8,
where again the data are missing. The decay from 1975 onwards is
also well matched (both in shape and amplitude) by the late decay of
pulse 70, which gives some support to this interpretation.

\subsection{Pulse energies}

\label{sec:Energies}

A further way to compare the pulses identified above is to compute
their energies. Integrating the segments of the light curves
corresponding to pulses 70, 81, 88, 90 and 92, we find that they
contain B-band energies of $0.60$, $0.61$, $0.48$, $0.73$\ and
$0.49\ET{51}$~erg respectively. While these values are similar to
each other, we note that pulses 88 and 92 are incomplete, so their
energies are certainly underestimated.

A direct integration of the light curve may not be the best way to
compute the pulse energies, since it is likely that not all the
nuclear luminosity is variable. In other words, the nucleus may well
harbour both a variable and a non-variable component. A non-variable
component could be associated with residual starlight from the host
galaxy or to a source intrinsic to the AGN. In the latter case, the
source could be associated with a nuclear star cluster, with the
non-varying portion of the putative accretion-disk, or with diffuse
emission from the vicinity of the nucler power-house. The strength
of this background component is not directly derivable from the
data, but an upper limit to its luminosity is given by the minimum
luminosity in the light curve. Subtracting the minimum observed
luminosity from the pulse light curves we obtain that pulses 70, 81,
88, 90 and 92 enclose B-band energies of $0.36$, $0.33$, $0.35$,
$0.51$\ and $0.33\ET{51}$~erg respectively. Obviously, both these
values and those obtained without the background subtraction are
subjected to the uncertainties in the absolute luminosity scales.

If the pattern recurrence suggested by the present analysis is
confirmed by further observations, it will be important to
investigate its wavelength dependence, particularly in the UV and
X-ray regions, where most of the bolometric luminosity of AGN
emerges. The detailed evolution of the pulses in different energy
bands and the estimation of bolometric energies would provide
stringent constraints for possible physical scenarios for the origin
of the pulses. The length and sampling rates of the presently
available UV and X-ray light curves of NGC 4151 and NGC 5548 do not
allow an analysis as detailed as for the optical band, but there is
strong evidence that the variations in all these bands are
correlated (e.g., Clavel \et 1992, Edelson \et 1996), which
indicates that the global variability patterns observed in the
optical are also present in the UV and X-rays. We note, however,
that NGC 5548 is substantially more luminous that NGC 4151 in both
UV and X-rays (Perola \et 1986, Clavel \et 1992, Edelson, Krolik \&
Pike 1990, Green, McHardy \& Lehto 1993, Korista \et 1995, Edelson
\et 1996), which indicates that the similarity between absolute
luminosities and energies of the pulses identified in the optical
light curves of these two galaxies does not hold over a broder
frequency range.

\section{Cross-correlation analysis}

\label{sec:XCORREL}

The initial method employed to find the pattern coincidences
described above was a simple eye inspection of the light curves, a 
method which, though admitedly subjective, proved quite successfull.
In this section we assess the statistical significance of these
findings.

\subsection{Method}

Probably the simplest method to identify occurrences of a given
pattern in a light curve is to first isolate the pattern we are
looking for and then cross-correlate it with the whole light curve
shifting the pulse along the time axis. High values of the
correlation coefficient would then indicate the positions where the
pattern fits best, and the statistical significance of the
correlation can be easily assessed. To compute such
cross-correlations we first have to  consider that the light curves
are irregularly sampled. We explored three different techniques to
deal with this difficulty: {\it (i)} intepolation, {\it (ii)} data
windowing and {\it (iii)} phase-binning. 

Interpolation can be done in one of several ways: interpolating both
light curves, interpolating only in the pulse light curve at the
dates of the full light curve or vice-versa. Any of these
alternatives involves some `data invention', which is particularly
critical for NGC 4151 given the long gaps in its light curve. 

An alternative method is to, after shifting the pulse to a given
date, find the pulse points within a given time window (say, $\pm
3$\ days) around each of the dates of the full light curve. For each
date in the light curve we obtain a set of corresponding pulse
luminosities which can then be averaged up, possibly weighting by
their distance to the center of the window and/or their error-bars,
to compute the correlation coefficient. Light curve dates which do
not have a corresponding pulse point in the time-window, are
excluded from the computation of the correlation coefficient. This
method does not involve any interpolation, but it suffers from a
more subtle problem. As is evident from the data, the observations
are inhomogenously distributed in time. Clustering of the
observations around small segments of the light curve (as, for
instance, in the 1981.2--1981.7 period in the NGC 4151 light curve)
inevitably bias the correlation coefficient, in the sense that
instead of measuring the matching of the pulse as a whole it gives a
disproportionally large weight to the matching on short time scales.

A third method (`phase-binning') consists of, after shifting the
pulse to a given position in the light curve, binning both pulse and
light curve into $\Delta t_{bin}$-wide bins, and correlating the
mean values of the two data trains in each of the (non-empty) bins.
In this way we solve the problem of over-weighting the matching on
short time-scales without resorting to interpolation. The size of
the bins is dictated by the time-scale to which we want the matching
to be measured. As we have seen, there is evidence for pattern
matchings on scales of weeks to years, so an appropriate bin size
would be of the order of 10 days. The cross-correlations presented
below were performed with this method, but we emphasize that, though
we judge phase-binning to be superior to interpolation and
data-windowing, all three methods were tried and found to give
similar results. 

\subsection{Results}

\begin{figure}  
    \protect\centerline{
    \epsfxsize=8.6cm\epsffile[50 175 470 690]{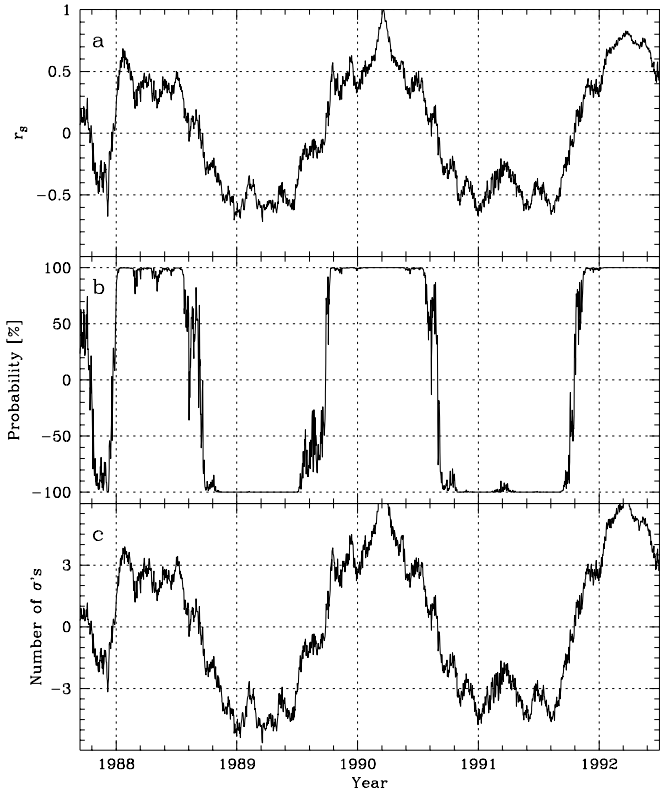}
     }
    \caption{
    Cross-correlation of pulse 90 with the full light curve of NGC
5548. (a) Spearman's rank correlation coefficient, (b) confidence
level of the correlation, and (c) number of standard deviations by
which the `sum of squared differences of ranks' deviates from the
expected value in the null hypothesis (no correlation). The sign
convention in panels b and c is that positive peaks correspond to
positive correlations and vice-versa. High peaks indicate possible
fitting dates for pulse 90.
    }
    \label{fig:XCORREL90}
\end{figure}

In Fig.~\ref{fig:XCORREL90} we plot the results of the
cross-correlation of pulse 90 with the whole light curve of NGC 5548
for a bin-size of 5 days. The top panel (a) shows Spearman's
non-parametric correlation (Press \et 1991) coefficient ($r_S$) as a
function of the date to which pulse 90 is shifted to. The
corresponding confidence level is plotted in panel b, where, for the
sake of clarity anti-correlations ($r_S < 0$) are represented by a
negative confidence level. The peak at 1990.2 corresponds to the
correlation of pulse 90 with itself, which naturally yields $r_S =
1$\ and 100\% confidence, while the peaks at 1988.1 and 1992.3
correspond to the correlations of pulse 90 with pulses 88 and 92
respectively. It is seen that these coincidences are highly
significant ($> 99\%$). This is better seen in panel c, where the we
plot the number of standard deviations by which the `sum of squared
differences of ranks' deviates from its expected value in the case
of no correlation (see Press \et 1991). The sign convention here is
as in panel b. Fig.~\ref{fig:XCORREL90}c shows that the correlation
of pulse 90 with pulse 88 is significant to a $4 \sigma$\ level,
while pulses 90 and 92 correlate even better ($6 \sigma$). The
mathematical analysis thus corroborates our visual identification of
pulse matchings in the light curve of NGC 5548.

Notice that the 1988.1, 1990.2 and 1992.3 peaks are surrounded by
adjacent troffs, where the anti-correlations are as significant as
in the positive peaks. Strong anti-correlations occur whenever the
pulse is shifted by $\sim \pm$\ half its duration from one of its
matching dates, i.e.\ as it moves `off-phase' by 1/2. The situation
here is analogous to what would happen if we correlated one cycle of
a sine wave with a long sinusoidal light curve: Positive peaks would
indicate matching phases and negative peaks would correspond to
locations were the pulse phase is off-set by $180^\circ$\ with
respect to the light curve.

\begin{figure}  
    \protect\centerline{
    \epsfxsize=8.6cm\epsffile[50 175 470 690]{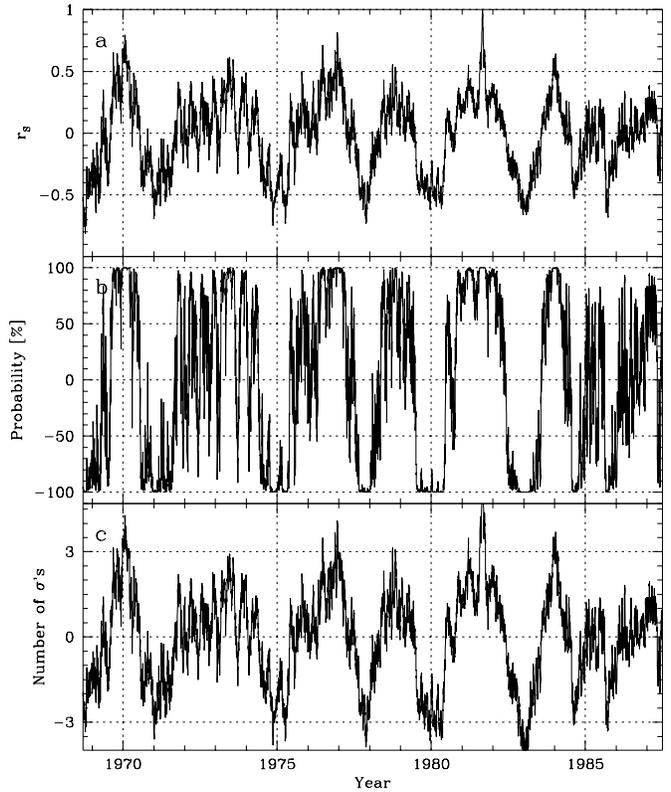}
     }
    \caption{
    As Fig.\ 4, but for the cross-correlation of pulse 81 with the
light curve of NGC 4151.
    }
    \label{fig:XCORREL81}
\end{figure}

Fig.~\ref{fig:XCORREL81} shows the cross-correlation of pulse 81
with the full light curve of NGC 4151 for $\Delta t_{bin} =
15$~days. The noisier appearance of this plot compared to 
Fig.~\ref{fig:XCORREL90} is mainly due to the worse sampling rate in
NGC 4151 (an average of 1 observation every 10 days, compared to 1
every 3 days for NGC 5548). Still, significant peaks are found at
the following dates 1970 ($4\sigma$), 1977 ($4\sigma$),  1978.8
($3\sigma$) and 1984 ($3.5 \sigma$), while the maximum at 1981.7
marks the correlation of pulse 81 with itself. Note that, with the
exception of 1984, these dates are very similar to those identified
visually (Section~\ref{sec:Other_Pulses}). What the
cross-correlation is detecting in the 1984 peak is the long term
matching of pulse 81 with the gradual increase in luminosity during
the 1984--1985.5 period and the subsequent decrease until 1986.6. 
Though the long term agreement is reasonable, the evidence for
matching on shorter time-scales is not compelling, as indeed is the
case for the 1977 and 1978.8 events. The broad hump in the
cross-correlation between 1972 and 1974 corresponds to the region
where, as argued above, a superposition of two pulses might have
occurred. The significance of the hump is not large ($< 3\sigma$),
but this is not surprising, since a single pulse can only partially
match the combination of two events. The most significant peak in
Fig.~\ref{fig:XCORREL81} (that at 1970) corresponds to the matching
between pulses 70 and 81 discussed in Section~\ref{sec:70X81}. On
the whole, it is reasuring to note that the highest peak in the
cross-correlation analysis corresponds to the most interesting
matching found in our visual analysis of NGC 4151,  and that other
peaks also coincide with visually estimated possible matchings
dates.

Cross-correlations as those performed above but using Pearson's
linear correlation-coefficient instead of the (more robust)
rank-order coefficient yield nearly identical results. We have also
correlated pulses 92 and 88 with the full NGC 5548 light curve, and
pulse 70 with NGC 4151. Not surpringly (given the similarity of the
pulses), very similar results were obtained. Finally, we remark that
the cross-correlations presented above are not strongly influenced
by the bin-size chosen. The main effect of $\Delta t_{bin}$\ is to
determine the smoothness of the cross-correlation functions (small
bins leading to `noiser' curves and vice-versa). The significance of
the peaks is only marginally affected by the bin-size, while their
locations remain unchanged. The smaller bin-size used of NGC 5548
reflects the much better sampling of the AGN Watch light curve
compared to that of NGC 4151. This better sampling allows us to
measure the pattern matching in more detail than for NGC 4151.

To conclude, the cross-correlation analysis confirmed the pulse
identifications of Section~\ref{sec:PULSES}. A much more direct and
stringent test of the hypothesis that variability patterns such as
those proposed here occur recurrently in type 1 Seyfert nuclei will
soon be provided by the release of the latest AGN Watch data for NGC
5548. Meanwhile, a further test of the similarities between pulses
88, 90 and 92 could be carried out comparing their spectral
properties (colours, line profiles, line ratios), something which
can be done with the currently available data.

\section{Discussion}

\label{sec:DISCUSSION}

The basic temporal structure of pulses 70, 81, 88, 90 and 92
consists of a short, not very energetic burst (subindex 1 in 
Fig.~\ref{fig:Pulses}b) followed by a $\sim 2$\ years long powerful
burst which shows a global decay with secondary peaks associated
with $\sim$\ month-scale oscillations (subindices 2--4 in
Fig.~\ref{fig:Pulses}b). This is obviously a simplified description
of the structure of the pulses, but it accounts for the most
energetic events in the pulse evolution. High frequency variations
are also present but these are usually of low amplitude (Lyutyi \et
1989, Dultzin-Hacyan \et 1992, Edelson \et 1994, 1996, Gopal-Krishna
\et 1995). 

Despite their similarities, it is clear that the pulses are not
identical. This, however, is not surprising, since an exact
repetition of a given pattern would only occur if the physical
conditions under which they are generated are identical, an extreme
requirement in any physical scenario. The fact that the observed
pulses are so similar suggests that their fundamental parameters are
similar for all the pulses in both NGC 4151 and NGC 5548. It is
interesting to note a similar patterns of variations has been
observed in the $H\alpha$\ light curve of NGC 1566 (Alloin \et
1986), but the time-scale of the slow component there seems to be
longer.

\subsection{Origin of the pulses}

If variability does follow a certain characteristic pattern, the
next question is what causes it. In this section we discuss possible
interpretations in the context of current AGN theories.

The conventional wisdom is that AGN variability is the result of
instabilities in an accretion disk around a supermassive black-hole
(e.g.\ Rees 1984, Abramowicz 1991). An important question in this
respect is whether such instabilities should produce well defined
patterns at all---theorists are strongly encouraged to address this
point. A variability pattern could perhaps be associated with the
evolution of an instability as it spirals  inwards to be finally
accretted by the hole, but this interpretation has to await detailed
modelling.

Recurrent patterns find a natural interpretation in models which
attribute the luminosity variations due to stochastic occurences of
same basic physical process. The best studied example of a model
fitting this general scenario is the so called starburst model for
AGN, which attributes the variations to supernova explosions and
their associated compact remnants in a massive young star cluster
(Terlevich \et 1992, 1995, Aretxaga \& Terlevich 1994). The starting
shot would correspond to the SN flash while the subsequent long
flare and its oscillations are explained in terms of the evolution
of the corresponding compact SN remnant (see comparison of pulse 70
with SNe light curves in Fig.~1 of Aretxaga \& Terlevich 1994). In
fact, hydrodynamical simulations of SN remnants evolving in high
density media produce light curves which resemble those in
Fig.~\ref{fig:Pulses} (Terlevich \et 1995, Plewa 1995). This model
makes specific predictions about the spectral properties and the
different lags between continuum and line variations for different
flares. In particular, if the identification of the starting shots
as SN flashes is correct the nuclear spectra should show spectral
signatures of type II SNe at the corresponding dates.

Another proccess which might promote variability is the tidal
disruption of stars by the gravitational field of a black-hole 
(Rees 1988, Evans \& Kochanek 1989, Eracleous, Livio \& Binnette
1995),  an idea recently discussed in connection with the $H\alpha$\
variations of LINERs such as NGC 1097 (Storchi-Bergmann \et 1995)
and M81 (Bower \et 1996). Intuitively one would expect that
different disruption events would go through similar evolutionary
phases, perhaps leading to a recurrent pattern of variations.

This discussion illustrates the importance of establishing whether
AGN variability follow well defined, recurrent patterns. Empirical
confirmation of a pattern would pose a strong constraint on
theoretical models, challenging them to match the observations
quantitatively. On the other hand, the absence of recurrent patterns
could be used to argue against models where the variations are
explained in terms of stochastic occurrences of the same type of
event.

\subsection{Connection with QSO variability}

As fundamental as the question of whether AGN variability follows
well defined patterns is the question of whether the variations are
due to a Poissonian (`Christmas-tree') type of process. This latter
point has been extensively discussed in relation to the variability
of QSOs. In a Poissonian process variability is the result of the
random superposition of independent pulses. Such a process would
result in a `$1 / \sqrt{N}$' anti-correlation between the net rms
variability and the mean luminosity. Observational studies of the
variability-luminosity relationship in QSOs indicate that the slope
of the anti-correlation is not as steep as $-1/2$\ (Pica \& Smith
1983, Hook \et 1994, Paltani \& Courvoisier 1994, Cristiani \et
1996). Such studies are however plagued by photometric and sampling
uncertainties as well as redshift effects. When these effects are
taken into account the data is found to be consistent with the
Poissonian model (Cid Fernandes, Aretxaga \& Terlevich 1996,
Aretxaga, Cid Fernandes \& Terlevich 1997, Cid Fernandes 1995).
Furthemore, a $1 / \sqrt{N}$\ law only applies if the energies and
time-scales of the pulses do not depend on the QSO luminosity (Cid
Fernandes, Aretxaga \& Terlevich 1996). Though far from being
confirmed, a Poissonian nature for the variations observed in QSOs
is by no means ruled out by the currently available data.

\begin{figure}  
    \protect\centerline{
    \epsfxsize=8.4cm\epsffile[40 170 565 500]{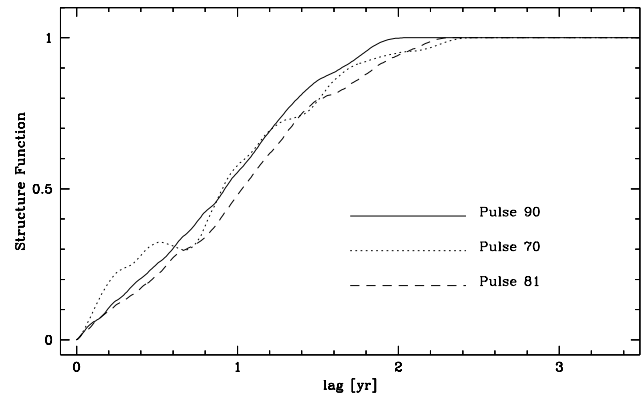}
     }
    \caption{
    Structure Functions of pulses 70, 81 and 90, normalized to their
assymptotic limt.
     }
    \label{fig:SF}
\end{figure}

If QSO light curves are indeed the result of the random
superposition of many pulses like those identified in the light
curves of NGC 4151 and NGC 5548, then their Structure Function (SF)
should be equal to the SF of the individual pulses. The SF at `lag'
$\tau$\ is defined as the mean value of $[L(t+\tau) - L(\tau)]^2$\
over the light curve, being basically an `up-side-down' version of
the auto-correlation function. The SF is particularly usefull to
pick up variability time-scales. The {\it ensemble} SF of large QSO
samples (e.g.\ Hook \et 1994, Cristiani \et 1996) shows a rapid rise
for lags shorter than $\sim 2$--$3$~yr and then flattens to a
roughly constant level, indicating that the duration of the pulses
is $\sim 2$--$3$~yr. This is comparable to the durations of pulses
70, 81 and 90, whose SF is plotted in Fig.~\ref{fig:SF}.

Regarding the energy of the pulses, Cid Fernandes, Aretxaga \&
Terlevich (1996) and Aretxaga, Cid Fernandes \& Terlevich (1997)
have shown that if QSO variability results from a Poissonian process
then the B-band energy of the individual pulses must be of the order
of $10^{50}$--$10^{51}$~ergs. These values are in excellent
agreement with those derived in Section~\ref{sec:Energies} (see also
Aretxaga \& Terlevich 1993, 1994), strengthening the idea that such
pulses constitute a `unit' of variability in active galaxies.

\section{Conclusions}

\label{sec:CONCLUSIONS}

We have reported the apparent existence of a recurrent pattern in
the optical light curves of NGC 4151 and NGC 5548. This pattern
consists of a short, not very energetic burst followed by a long,
powerful one which shows a global decay with month-scale
oscillations. Possible interpretations in terms of accrettion disk
instabilities, supernovae and stellar disruptions were briefly
outlined. A possible connection with the variability properties of
QSOs was also discussed. We found that the energies and time-scales
of the pulses identified in NGC 4151 and NGC 5548 are consistent
with those derived from QSO light curves under the assumption that
QSO variability is the result of a Poissonian process.

The evidence, though suggestive, has to be taken with caution given
the difficulties associated with the irregular sampling in the
present data sets. Continuing monitoring of these two sources and
other low luminosity AGN will certainly be able to conclusively
establish whether the suggested pattern is real or just a rare
coincidence.


\vskip 3mm

{\bf Acknowledgments:} We are greatly indebted to Ton Snijders and the
late Michael Penston for the data on NGC 4151 and to Bradley Peterson
and the whole AGN Watch consortium for making the data on NGC 5548
available to us previous to publication. We thank Mike Irwin, Laerte
Sodr\'e\ and Guillermo Tenorio-Tagle for discussions. We are also
thankful to the anonymous referee for valuable suggestions. 
The hospitality of INAOE (Mexico), where part of this work was carried out,
was much appreciated. The work of
RCF was supported by CAPES (grant 417/90-5) and CNPq (3000867/95-6). IA
was supported by the EEC HCM fellowship ERBCHBICT941023.



\end{document}